\documentclass{aa}
\usepackage{graphicx}
\usepackage{bm}
\graphicspath{{./fig/}{./png/}}

\newcommand{\EQ}{\begin{equation}}
\newcommand{\EN}{\end{equation}}
\newcommand{\EQA}{\begin{eqnarray}}
\newcommand{\ENA}{\end{eqnarray}}

\newcommand{\EEq}[1]{Equation~(\ref{#1})}
\newcommand{\Eq}[1]{Equation~(\ref{#1})}

\newcommand{\Sec}[1]{Section~\ref{#1}}

\newcommand{\Fig}[1]{Figure~\ref{#1}}

\newcommand{\Figs}[2]{Figures~\ref{#1} and \ref{#2}}

\newcommand{\Tab}[1]{Table~\ref{#1}}

\newcommand{\bra}[1]{\langle #1\rangle}

{}
{}
{}

{}
{}
\newcommand{\meanEMF}{\overline{\mbox{\boldmath ${\cal E}$}}{}}{}
{}
{}
{}
{}
{}
\newcommand{\meanAA}{\overline{\mbox{\boldmath $A$}}{}}{}
\newcommand{\meanBB}{\overline{\mbox{\boldmath $B$}}{}}{}
{}
{}
{}
{}
{}
{}
{}
\newcommand{\meanJJ}{\overline{\mbox{\boldmath $J$}}{}}{}

{}
{}
{}

\newcommand{\meanB}{\overline{B}}

{}

{}
{}

\newcommand{\alphaK}{\alpha_{\rm K}}
\newcommand{\alphaM}{\alpha_{\rm M}}
\newcommand{\nullvector}{{\bf0}}

\newcommand{\uu}{\mbox{\boldmath $u$} {}}
\newcommand{\UU}{\mbox{\boldmath $U$} {}}

\newcommand{\BB}{\mbox{\boldmath $B$} {}}

\newcommand{\JJ}{\mbox{\boldmath $J$} {}}

\newcommand{\AAA}{\mbox{\boldmath $A$} {}}

\newcommand{\ff}{\mbox{\boldmath $f$} {}}

\newcommand{\FF}{\mbox{\boldmath $F$} {}}

\newcommand{\nab}{\mbox{\boldmath $\nabla$} {}}

\newcommand{\SSSS}{\mbox{\boldmath ${\sf S}$} {}}

\newcommand{\erf}{{\rm erf}}

\newcommand{\DD}{{\rm D} {}}

\newcommand{\dd}{{\rm d} {}}

\newcommand{\const}{{\rm const}  {}}

\def\Pm{\mbox{\rm Pr}_M}
\def\Rm{\mbox{\rm Re}_M}

\def\Rey{\mbox{\rm Re}}

\def\cs{c_{\rm s}}

\def\kf{k_{\rm f}}

\def\Brms{B_{\rm rms}}

\def\urms{u_{\rm rms}}

\def\etat{\eta_{\rm t}}
\def\etatz{\eta_{\rm t0}}
\def\etaTz{\eta_{\rm T0}}

\def\Beq{B_{\rm eq}}

\def\half{{\textstyle{1\over2}}}

\def\onethird{{\textstyle{1\over3}}}

\newcommand{\yapj}[3]{ #1, {ApJ,} {#2}, #3}

\newcommand{\yapjl}[3]{ #1, {ApJ,} {#2}, #3}

\newcommand{\yan}[3]{ #1, {Astron.\ Nachr.,} {#2}, #3}

\newcommand{\yana}[3]{ #1, {A\&A,} {#2}, #3}

\newcommand{\sgafd}[1]{ #1, {Geophys.\ Astrophys.\ Fluid Dyn.,} submitted}

\newcommand{\yjfm}[3]{ #1, {J.\ Fluid Mech.,} {#2}, #3}

\newcommand{\yprl}[3]{ #1, {Phys.\ Rev.\ Lett.,} {#2}, #3}

\newcommand{\ypre}[3]{ #1, {Phys.\ Rev.\ E,} {#2}, #3}

\newcommand{\yjour}[4]{ #1, {#2}, {#3}, #4}

\newcommand{\ybook}[3]{ #1, {#2} (#3)}

\newcommand{\sapj}[1]{ #1, {ApJ}, submitted}

\begin{document}

\titlerunning{Surface appearance of dynamo-generated fields}
\authorrunning{Warnecke \& Brandenburg}

\title{Surface appearance of dynamo-generated large-scale fields}
\author{J. Warnecke\inst{1,2} \and A. Brandenburg\inst{1,2}}
\institute{Nordita, AlbaNova University Center, Roslagstullsbacken 23,
SE-10691 Stockholm, Sweden
\and Department of Astronomy, AlbaNova University Center,
Stockholm University, SE 10691 Stockholm, Sweden}

\date{\today,~ $ $Revision: 1.88 $ $}
\abstract{}{%
Twisted magnetic fields are frequently seen to emerge above the visible
surface of the Sun.
This emergence is usually associated with the rise of buoyant magnetic
flux structures.
Here we ask how magnetic fields from a turbulent large-scale dynamo
appear above the surface if there is no magnetic buoyancy.
}{%
The computational domain is split into two parts.
In the lower part, which we refer to as the turbulence zone,
the flow is driven by an assumed helical forcing
function leading to dynamo action.
Above this region, which we refer to as the exterior,
a nearly force-free magnetic field is computed
at each time step using the stress-and-relax method.
}{%
Twisted arcade-like field structures are found to emerge
in the exterior above the turbulence zone.
Strong current sheets tend to form above the neutral
line, where the vertical field component vanishes.
Time series of the magnetic field structure show recurrent plasmoid 
ejections.
The degree to which the exterior field is force free is estimated
as the ratio of the dot product of current density and magnetic field
strength to their respective rms values.
This ratio reaches values of up to 95\% in the exterior.
A weak outward flow is driven by the residual Lorentz force.
}{}

\keywords{magnetohydrodynamics (MHD) -- turbulence --
Sun: dynamo -- stars: magnetic fields
}

\maketitle

\section{Introduction}

The magnetic field at the visible surface of the Sun is known to take
the form of bipolar regions.
Above these magnetic concentrations the field continues in an arch-like fashion.
These formations appear usually as twisted loop-like structures.
These loops can be thought of as a continuation of more concentrated
flux ropes in the bulk of the solar convection zone.
However, this interpretation is problematic because we cannot be certain
that the magnetic field in the Sun is generated in the form of flux ropes.
Indeed, simulations of successful large-scale dynamos suggest that
concentrated tube-like structures are more typical of the early kinematic
stage, but in the final nonlinear stage the field becomes more space-filling
(Brandenburg 2005, K\"apyl\"a et al.\ 2008).
The idea that the dynamics of such tubes is governed by magnetic
buoyancy is problematic too, because the solar convection zone is strongly
stratified with concentrated downdrafts and broader upwellings.
This leads to efficient downward pumping of magnetic field toward
the bottom of the convection zone (Nordlund et al.\ 1992; Tobias et al.\ 1998).
This downward pumping is generally found to dominate over magnetic buoyancy.
The question then emerges whether magnetic buoyancy can still be invoked
as the main mechanism for causing magnetic flux emergence at the solar
surface.
Another possible mechanism for the emergence of magnetic field at the
solar surface might simply be the relaxation of a strongly twisted
magnetic field in the bulk of the convection zone.
Twisted magnetic fields are produced by a large-scale dynamo mechanism
that is generally believed to be the motor of solar activity (Parker 1979).
One such dynamo mechanism is the $\alpha$ effect that produces a large-scale
poloidal magnetic field from a 
toroidal
one.
However, this mechanism is known to produce magnetic fields of
opposite helicity (Seehafer 1996; Ji 1999).
This magnetic helicity of opposite sign is an unwanted by-product,
because it quenches the dynamo effect (Pouquet et al.\ 1976).
A commonly discussed remedy is therefore to allow the helicity of
small-scale field to leave the domain, possibly in the form of coronal
mass ejections (Blackman \& Brandenburg 2003).

In order to study the emergence of helical magnetic fields from a
dynamo, we consider a model that combines a direct simulation of a
turbulent large-scale dynamo with a simple treatment of the evolution
of nearly force-free magnetic fields above the surface of the dynamo.
An additional benefit of such a study is that it alleviates the need
for adopting a boundary condition for the magnetic field at the top of
the dynamo region.
This is important, because it is known that the properties of the generated
large-scale magnetic field strongly depend on boundary conditions.
A common choice for the outer boundary condition is to assume that the magnetic
field can be matched smoothly to a potential field.
Such a condition is relatively easily implemented in calculations employing
spherical harmonic functions (see, e.g., Krause \& R\"adler 1980).
A more realistic boundary condition might be an extrapolation to a force-free
magnetic field where the Lorentz force vanishes in the exterior.
This means that the current density is proportional to the local magnetic
field, but the constant of proportionality depends generally on the
magnetic field itself.
This renders the magnetic boundary condition nonlinear,
which is therefore  not easy to implement.
Moreover, a perfectly force-free magnetic field is not completely
realistic either.
Instead, we know that above the solar surface, magnetic
fields drive flares and coronal mass ejections through the Lorentz force.
A more comprehensive approach would be to include in the
calculations the exterior regions above the solar or stellar surface.
This can be computationally prohibitive and a realistic treatment
of that region may not even be necessary.
It may therefore make sense to look for simplifying alternatives.
One possibility is therefore to attempt an iteration toward a nearly
force-free magnetic field such that the field can deviate from a force-free
state locally in regions where the field cannot easily be made force-free.
This could be done by solving the induction equation with an additional
ambipolar diffusion term, which implies the presence of an effective
velocity correction proportional to the local Lorentz force.
This is sometimes called the magneto-frictional method and has been
introduced by Yang et al.\ (1986) and Klimchuk \& Sturrock (1992).
In this approach the electromotive force attains not only a term in
the direction of $\BB$, but also a term in the direction of $\JJ$
(Brandenburg \& Zweibel 1994).
The latter corresponds to a diffusion term, which explains the
diffusive aspects of this effect.
However, the resulting ambipolar diffusivity coefficient is proportional to
$\BB^2$ and can locally become so large that the computational time step
becomes significantly reduced.
This is a typical problem of parabolic equations.
A better method is therefore to turn the problem into a
hyperbolic one and to solve an additional evolution equation
for the velocity correction where the driving force is the Lorentz force.
This approach is sometimes called the ``force-free model'' (FFM), even though
the field in this model is never exactly force-free anywhere
(Miki\'c et al.\ 1988; Ortolani \& Schnack 1993).
In the context of force-free magnetic field extrapolations this method is
also known as the stress-and-relax method (Valori et al.\ 2005).

\section{Equations for the Force-Free Model}
\label{FFM}

The equation for the velocity correction in the Force-Free Model (FFM) is similar to the
usual momentum equation, except that there is no pressure, gravity, or other
driving forces on the right-hand side.
Thus, we just have
\begin{equation}
{\DD\UU\over\DD t}=\JJ\times\BB/\rho+\FF_{\rm visc},
\label{DUDtext}
\end{equation}
where $\JJ\times\BB$ is the Lorentz force,
$\JJ=\nab\times\BB/\mu_0$ is the current density,
$\mu_0$ is the vacuum permeability,
$\FF_{\rm visc}$ is the viscous force,
and $\rho$ is here treated as a constant
corresponding to a prescribed density.
\EEq{DUDtext} is solved together with the induction equation.
In order to preserve $\nab\cdot\BB=0$, we write $\BB=\nab\times\AAA$ in terms
of the vector potential $\AAA$ and solve the induction equation in the form
\begin{equation}
{\partial\AAA\over\partial t}=\UU\times\BB+\eta\nabla^2\AAA,
\end{equation}
where we have adopted the so-called resistive gauge in which the electrostatic
potential is equal to $-\eta\nab\cdot\AAA$ and the magnetic diffusivity $\eta$
is assumed constant. 
The value of $\eta$ will be given in terms of the magnetic Reynolds number,
whose value will be specified below.
No continuity equation for $\rho$ is solved in this part of the domain, because
there is no pressure gradient in the momentum equation.

In the following we couple such a model for the magnetic field above the
solar photosphere to a simulation of a large-scale dynamo.
In order to keep matters simple, we restrict ourselves to the
case of an isothermal equation of state with constant sound speed $\cs$.
Our goal is then to analyze the appearance of the resulting magnetic
field above the surface of the dynamo and to study also the effects on
the dynamo itself.

\section{The model}
\label{TheModel}

The idea is to combine the evolution equations for the dynamo interior
with those of the region above by simply turning off those
terms that are not to be included in the upper part of the domain.
We do this with error function profiles of the form
\begin{equation}
\theta_w(z)=\half\left(1-\erf{z\over w}\right),
\end{equation}
where $w$ is the width of the transition.
Thus, the momentum equation is assumed to take the form
\begin{equation}
{\DD\UU\over\DD t}=\theta_w(z)\left(-\nab h+\ff\right)
+\JJ\times\BB/\rho+\FF_{\rm visc},
\end{equation}
where $\FF_{\rm visc}=\rho^{-1}\nab\cdot(2\rho\nu\SSSS)$ is the viscous force,
${\mathsf S}_{ij}=\half(U_{i,j}+U_{j,i})-\onethird\delta_{ij}\nab\cdot\UU$
is the traceless rate-of-strain tensor, commas denote partial
differentiation, $h=\cs^2\ln\rho$ is the specific pseudo-enthalpy,
$\cs=\const$ is the isothermal sound speed,
and $\ff$ is a forcing function that drives turbulence in the interior.
The pseudo-enthalpy term emerges from the fact that for an isothermal
equation of state the pressure is given by $p=\cs^2\rho$, so the pressure
gradient force is given by $\rho^{-1}\nab p=\cs^2\nab\ln\rho=\nab h$.
The continuity equation can either be written in terms of $h$
\begin{equation}
{\DD h\over\DD t}=-c_s^2\theta_w(z)\nab\cdot\UU,
\label{dhdt}
\end{equation}
where we have inserted the $\theta_w(z)$ factor to terminate the
evolution of $h$ in the exterior, or in terms of $\rho$,
\begin{equation}
{\partial\rho\over\partial t}=-\nab\cdot\left[\theta_w(z)\,\rho\UU\right],
\label{drhodt}
\end{equation}
which serves the same purpose, but also preserves total mass.
Most of the runs presented below are carried out using \Eq{dhdt},
but comparisons using \Eq{drhodt} resulted in rather similar behavior.

The forcing function consists of random plane helical transversal
waves with wavenumbers that lie in a band around an average forcing
wavenumber $\kf$.
These waves are maximally helical with $\nab\times\ff\approx\kf\ff$,
so the helicity is positive.
This type of forcing was also adopted in Brandenburg (2001) and
many other recent papers.
The profile function $\theta_w(z)$ in front of the forcing term restricts
the occurrence of turbulence mostly to the dynamo region, $z<0$.
The forcing amplitude is chosen such that the rms velocity in this
region, $\urms$, is about 4\% of the sound speed.

We adopt non-dimensional units by measuring density in units of the
initially constant density $\rho_0$, velocity in units of $\cs$,
and length in units of $k_1^{-1}$, where $k_1=2\pi/L_x$ is the minimal
wavenumber in the $x$ direction with an extent $L_x$.
In most of the cases reported below,
the vertical extent is $L_{z1}\leq z\leq L_{z2}$
with $L_{z1}=-L_x/3$ and $L_{z2}=2L_x/3$.
In a few cases we shall consider larger domains where
the domain is two or four times larger in the $z$ direction
than in the horizontal directions.
The extent of the domain in the $y$ direction is $L_y=L_x$.
We adopt periodic boundary conditions in the $x$ and $y$ directions.
For the velocity we employ stress-free boundary conditions at top
and bottom, i.e.\
\begin{equation}
U_{x,z}=U_{y,z}=U_z=0\quad\mbox{on $z=L_{z1}$, $L_{z2}$},
\end{equation}
for the magnetic field we adopt perfect conductor boundary conditions
at the bottom, which corresponds to
\begin{equation}
A_{x}=A_{y}=A_{z,z}=0\quad\mbox{on $z=L_{z1}$},
\end{equation}
and vertical-field or pseudo-vacuum conditions at the top, i.e.,
\begin{equation}
A_{x,z}=A_{y,z}=A_{z}=0\quad\mbox{on $z=L_{z2}$}.
\end{equation}
Note that no mass is allowed to escape at the top.
Although this restriction does not seem to affect the results of
our simulations significantly, it might be useful to adopt in future
applications outflow boundary conditions instead.

Our model is characterized by several dimensionless parameters.
Of particular importance is the magnetic Reynolds number,
\EQ
\Rm=\urms/\eta\kf,
\EN
where $\kf$ is the wavenumber of the energy-carrying eddies.
The ratio of viscosity to magnetic diffusivity is the magnetic Prandtl
number, $\Pm=\nu/\eta$.
In all our simulations we use $\Pm=1$.
The typical forcing wavenumber, expressed in units of the box wavenumber,
$\kf/k_1$, is another important input parameter.
In our simulations this value is 10.
For the profile functions we take a transition width $w$ with
$k_1 w=0.1$ in most of the runs, but in some cases it is 0.2.
The magnetic field is expressed in terms of the equipartition value,
$\Beq$, where $\Beq^2=\mu_0\bra{\rho\uu^2}$, and the ave\-rage is
taken over the turbulence zone.
We measure time in non-dimensional units $\tau = t \urms\kf$, which is
the time normalized to the eddy turnover time of the turbulence.
As initial condition we choose a hydrostatic state, $\UU=\nullvector$,
with constant density $\rho=\rho_0$, and the components of the magnetic
vector potential are random white noise in space with Gaussian statistics
and low amplitude ($10^{-4}$ below equipartition).

In this paper we present both direct numerical simulations and
mean-field calculations.
In both cases we use the
{\sc Pencil Code}\footnote{\texttt{http://pencil-code.googlecode.com}},
which is a modular high-order code (sixth order in space and third-order
in time) for solving a large range of different partial differential
equations.

\section{Results}

\begin{figure}[t!]\begin{center}
\includegraphics[width=\columnwidth]{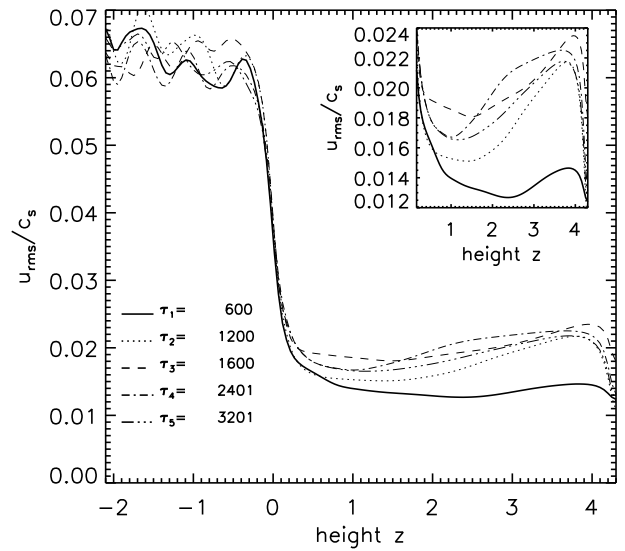}
\end{center}\caption[]{
Vertical dependence of the rms velocity at different times.
Note the drop of $\urms(z)$ from the turbulence zone to the
exterior by a factor of about 3--5
The inset shows $\urms(z)$ in the exterior at different times.
}\label{fig_urms}
\end{figure}

We begin by considering first hydrodynamic and hydromagnetic
properties of the model.
In the turbulence zone the velocity reaches quickly
a statistically steady value, while in the exterior it takes
about 1000 turnover times before a statistically steady state is reached.
This is seen in \Fig{fig_urms}, where we show $\urms(z)$ at different times.
In the following we discuss the properties of the magnetic field that is
generated by the turbulence.

\subsection{Dynamo saturation}

\begin{figure}[t!]\begin{center}
\includegraphics[width=\columnwidth]{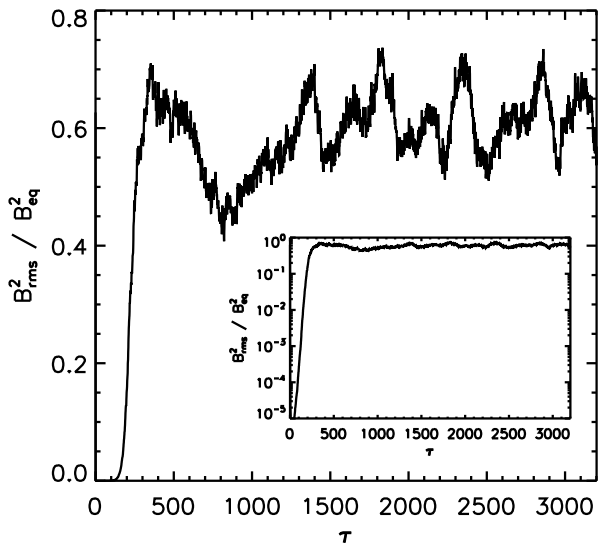}
\end{center}\caption[]{
Initial exponential growth and subsequent saturation behavior of
the magnetic field in the interior for forced turbulence with dynamo action.
}\label{fig_brms}
\end{figure}

Dynamo action is possible when $\Rm$ reaches a critical
value $\Rm^{\rm{crit}}$ that is about $0.5$ in our case.
The situation is only slightly modified compared with dynamo saturation
in a periodic domain.
For not too large values of $\Rm$ the dynamo saturates relatively swiftly, but for
larger values of $\Rm$ the magnetic field strength may decline with increasing
value of $\Rm$ (Brandenburg \& Subramanian 2005).
An example of the saturation behavior is shown in \Fig{fig_brms}, where $\Rm\approx3.4$.
The initial saturation phase ($100\leq\tau\leq500$) is suggestive of the
resistively slow saturation found for periodic domains (Brandenburg 2001),
but then the field declines somewhat.
Such a decline is not normally seen in periodic domains, but is typical
of dynamo action in domains with open boundaries or an external halo
(see Fig.~5 of Hubbard \& Brandenburg 2010a).
The field strength is about $78\%$
of the equipartition field strength, $\Beq$.

\begin{figure}[t!]\begin{center}
\includegraphics[width=\columnwidth]{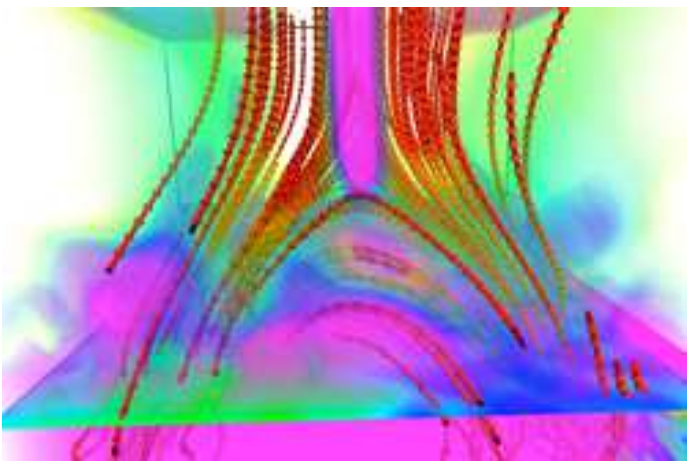}
\end{center}\caption[]{
Magnetic field structure in the dynamo exterior at $\tau$ = 1601.
Field lines are shown in red and the modulus of the current density
is shown in pink with semi-transparent opacity.
Note the formation of a vertical current sheet above the arcade.
}\label{fig_struc}
\end{figure}

\begin{figure}[t!]\begin{center}
\includegraphics[width=\columnwidth]{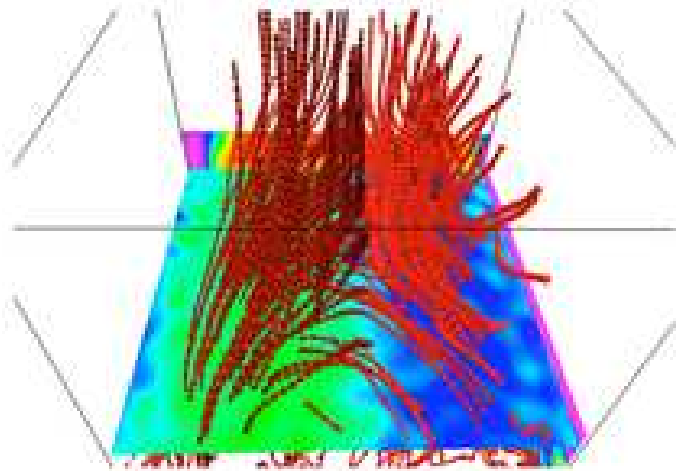}
\end{center}\caption[]{
Structure of magnetic field lines in the exterior,
together with a representation of the normal component of the field
at the interface at $z=0$ between turbulence zone and exterior at $\tau$ = 1601.
Green represents a positive and blue a negative value of $B_z$.}
\label{fig_curr}
\end{figure}

In all cases the magnetic field is strongest in the turbulence zone,
but it always shows a systematic variation in one of the two horizontal
directions.
It is a matter of chance whether this variation is in the $x$ or in the $y$
direction.
Comparison of different runs shows that both directions are about equally
possible (see below).
Also, the magnetic field pattern shows sometimes a slow horizontal
migration, but this too seems to be a matter of chance, as will be
discussed below.

\subsection{Arcade formation}

\begin{figure*}[t!]\begin{center}
\includegraphics[width=18cm]{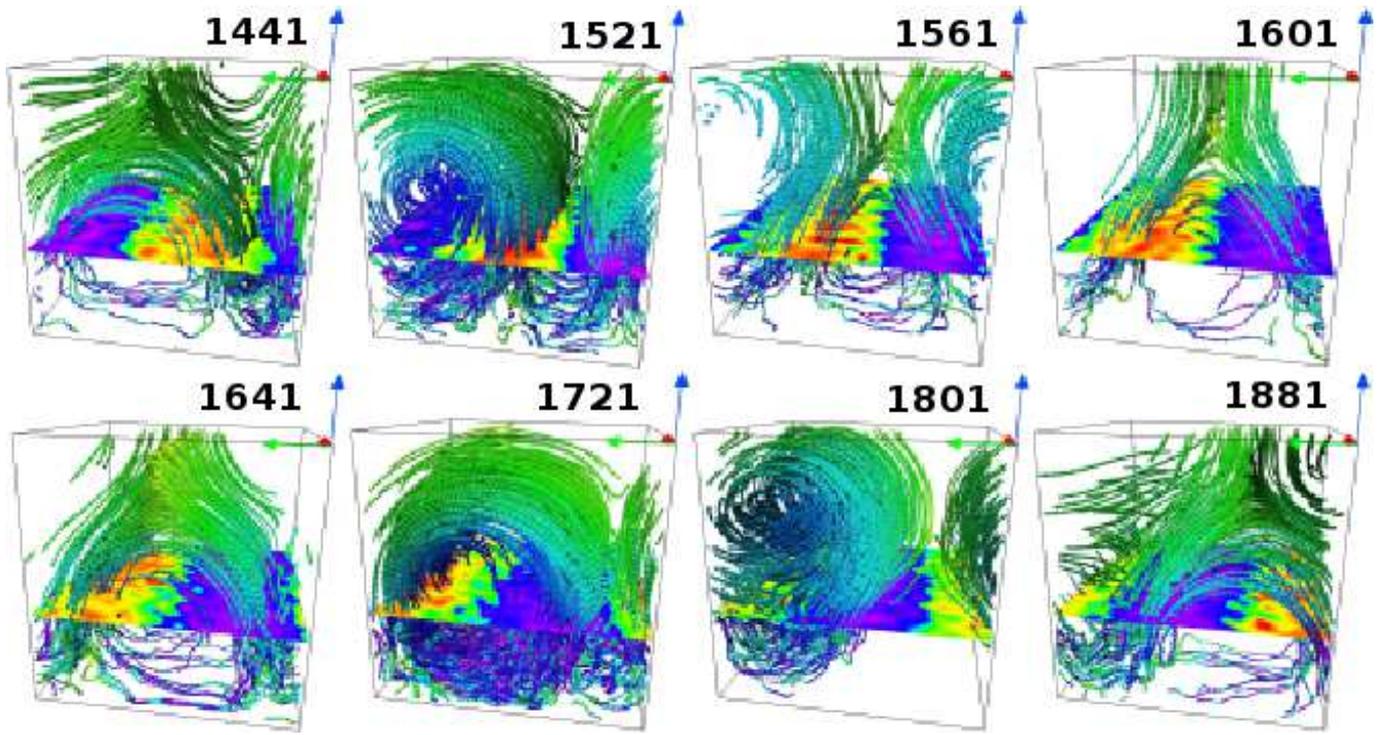}
\end{center}\caption[]{
Time series of arcade formation and decay.
Field lines are colored by their local field strength which increases
from pink to green.
The plane shows $B_z$ increasing from red (positive) to pink (negative).
The normalized time $\tau$ is giving in each panel.
}
\label{AI}
\end{figure*}

After some time the magnetic field extends well into
the exterior regions where it tends to produce an arcade-like structure,
as seen in \Figs{fig_struc}{fig_curr}.
The arcade opens up in the middle above the line where the vertical
field component vanishes at the surface.
This leads to the formation of anti-aligned field lines with a
current sheet in the middle; see \Figs{fig_struc}{fig_curr}.
The dynamical evolution
is seen clearly in a sequence of field line images in \Fig{AI}
where anti-aligned vertical field lines reconnect above
the neutral line and form a closed arch with plasmoid ejection above.
This arch then changes its connectivity at the foot points in the sideways
direction (here the $y$ direction), making the field lines bulge upward
to produce a new reconnection site with anti-aligned field lines some
distance above the surface.
Note that this sideways motion takes the form of a slowly propagating wave.
However, it is a matter of chance whether this wave propagates in the
positive or negative coordinate direction, as will be shown below
in \Sec{ForceFree}.

\begin{figure*}[t!]\begin{center}
\includegraphics[width=18cm]{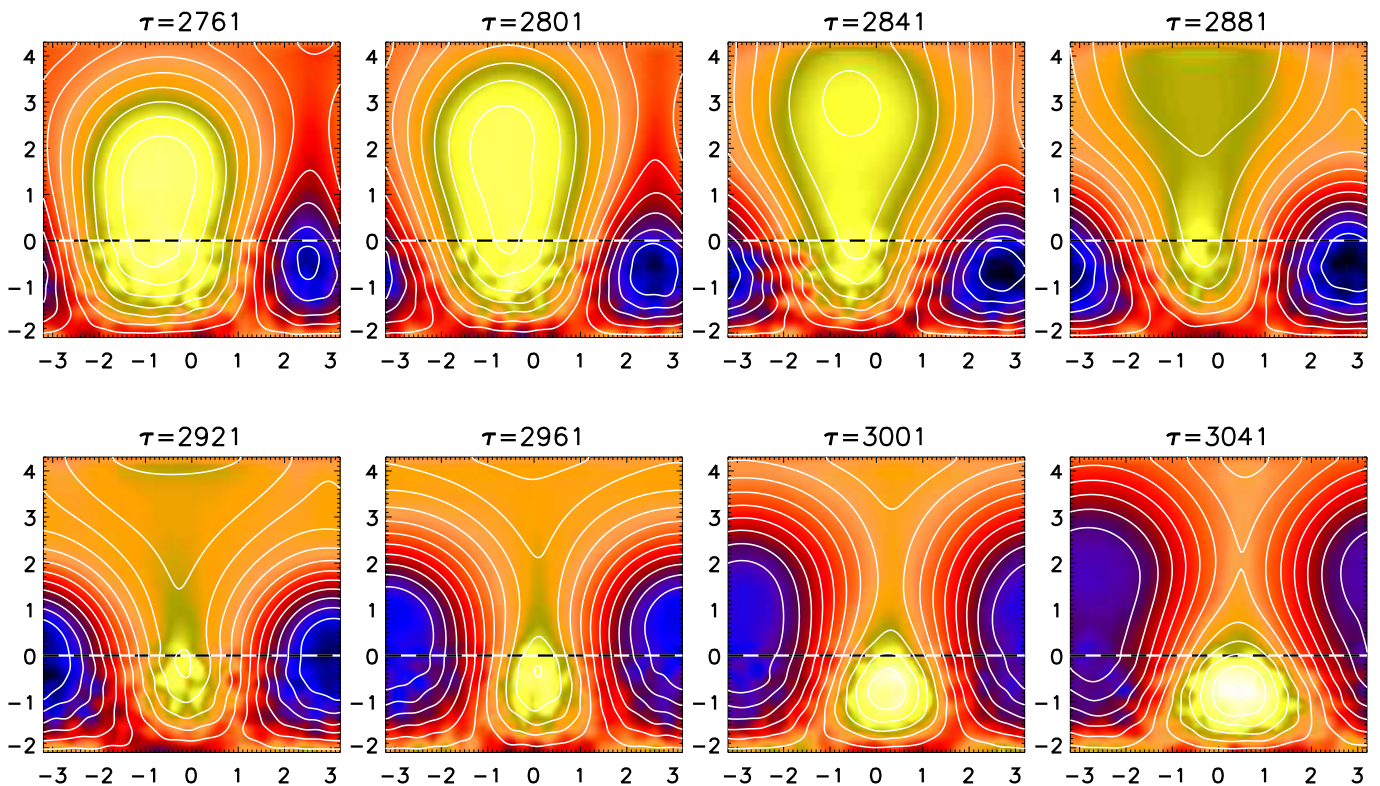}
\end{center}\caption[]{
Time series of the formation of a plasmoid ejection.
Contours of $\bra{A_x}_{x}$ are shown together with a
color-scale representation of $\bra{B_x}_x$;
dark blue stands for negative and red for positive values.
The contours of $\bra{A_x}_{x}$ correspond to field lines of $\bra{\BB}_x$
in the $yz$ plane.
The dotted horizontal lines show the location of the surface at $z=0$.
}
\label{aax}
\end{figure*}

Field line reconnection is best seen for two-dimensional magnetic fields,
because it is then possible to compute a flux function whose contours
correspond to field lines in the corresponding plane.
In the present case the magnetic field varies only little in the
$x$ direction, so it makes sense to visualize the field averaged in the
$x$ direction.
Since the averaging commutes with the curl operator, we can also average
the $x$ component of the magnetic vector potential, i.e.\ we compute
$\bra{A_x}_{x}$, where the second subscript indicates averaging along the
$x$ direction.
This function corresponds then to  the flux function of the magnetic field
in the $yz$ plane and averaged along the $x$ direction.
In \Fig{aax} we plot contours of $\bra{A_x}_{x}$, which correspond to
poloidal field lines of $\bra{\BB}_x$ in the $yz$ plane.
This figure shows clearly the recurrent reconnection events with
subsequent plasmoid ejection.
We also compare with a color/grey scale representation of the $x$ component
of the $x$-averaged magnetic field, $\bra{B_x}_x$.
Note that in the exterior the contours of $\bra{B_x}_x$ trace nearly
perfectly those of $\bra{A_x}_x$.
This suggests that the $x$-averaged magnetic field has nearly maximal
magnetic helicity in the exterior.
This is also in agreement with other indicators that will be considered below.

\subsection{Averaged field properties}

\begin{figure}[t!]\begin{center}
\includegraphics[width=\columnwidth]{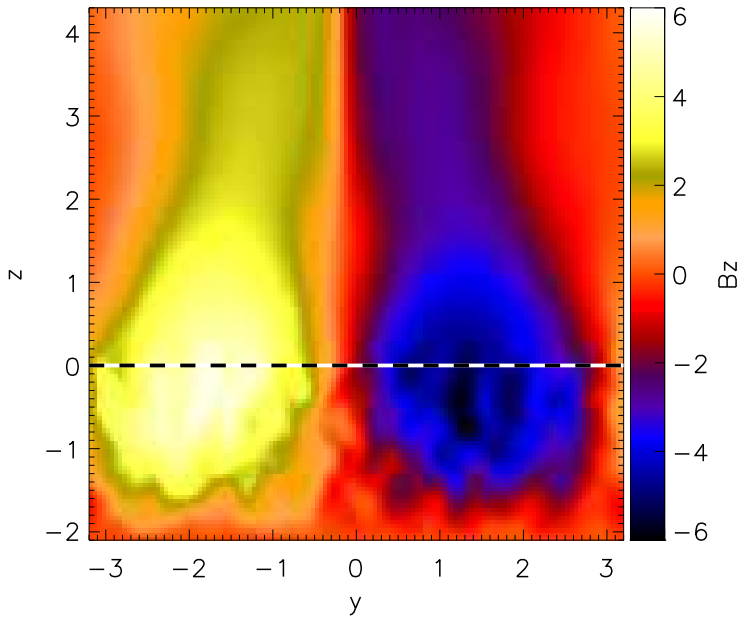}
\end{center}\caption[]{
Slice of $B_z$ through an arbitrarily chosen cross-section $x=\const$.
Note the periodicity with nearly sinusoidal variation in the $y$ direction,
and the more nearly monotonous fall-off in the $z$ direction.
}\label{cross-section}
\end{figure}

The magnetic field is largely confined to the turbulence zone where it
shows a periodic, nearly sinusoidal variation in the $y$ direction.
Away from the turbulence zone the field falls off, as can be seen
from the vertical slice shown in \Fig{cross-section}.
Near the top boundary, some components of the field become stronger again,
but this is probably an artifact of the vertical field condition employed
in this particular case.

\begin{figure}[t!]\begin{center}
\includegraphics[width=\columnwidth]{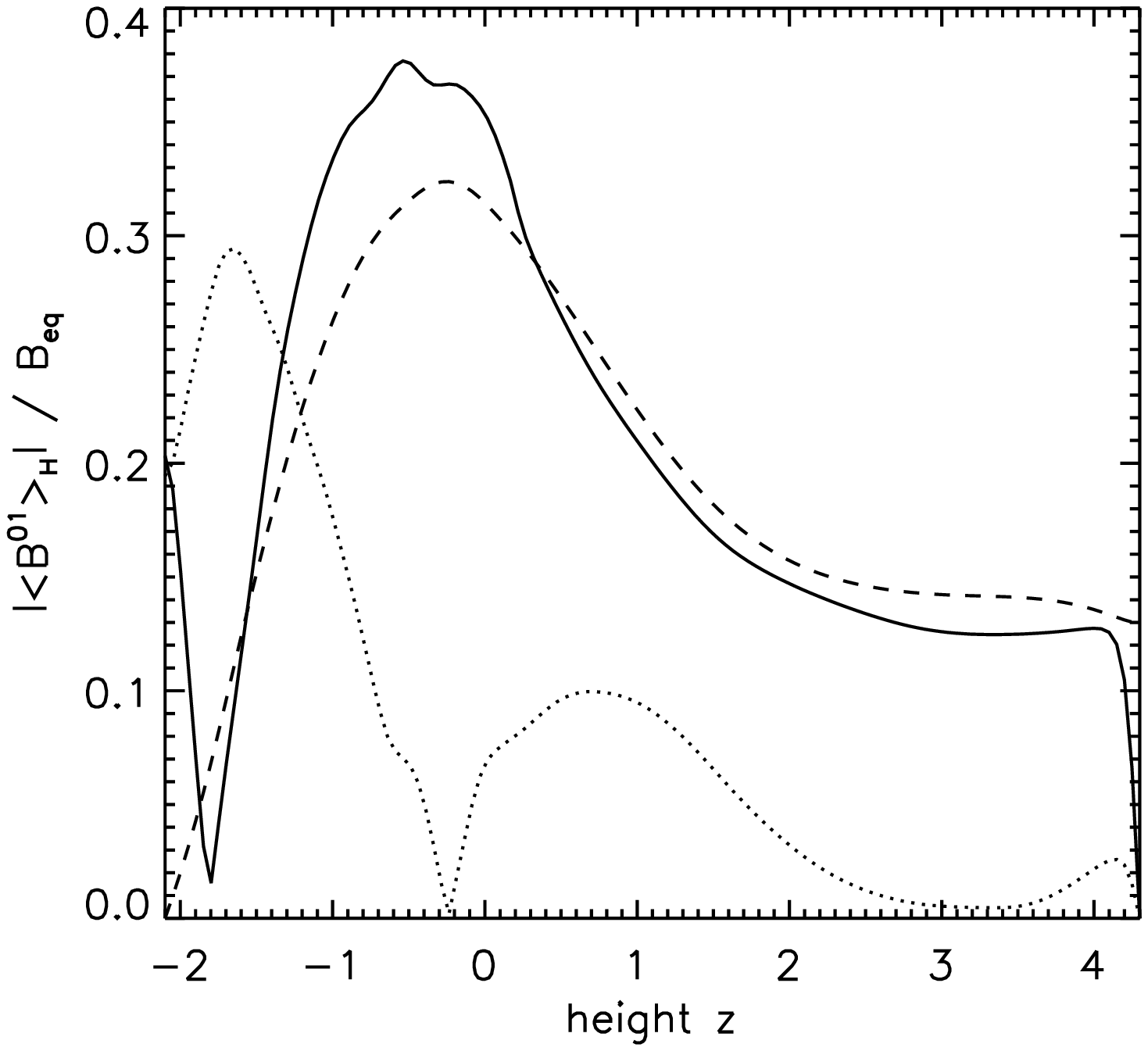}
\end{center}\caption[]{
Vertical dependence of the moduli of the components of $\meanB$,
as defined in \Eq{MFdef}.
}\label{fig_butt}
\end{figure}

In order to describe the vertical variation of the magnetic field in
an effective manner, it is appropriate to Fourier-decompose the field
in the two horizontal directions and to define a complex-valued
mean field as
\begin{equation}
\meanB_i^{lm}(z,t)=\!\int\!\!\int
{\dd x\over L_x}{\dd y\over L_y}
\BB(x,y,z,t)\,e^{2\pi i(lx/L_x+my/L_y)},
\label{MFdef}
\end{equation}
where superscripts $l$ and $m$ indicate a suitable Fourier mode.
In \Fig{fig_butt} we plot absolute values of the three components of
$\meanBB^{01}(z,t)$ as a function of $z$
for a time representing the final saturated state.
This figure shows quite clearly a relatively rapid decline of
$|\meanB_y|$ with $z$, while $|\meanB_x|$ and $|\meanB_z|$ level
off at values that are still about 40\% of that in the turbulence zone.
This suggests that our model is suitable to describe
the evolution of magnetic fields in the dynamo exterior.
Earlier simulations of coronal loops and coronal heating
(Gudiksen \& Nordlund 2002, 2005; Peter et al.\ 2004) demonstrate
that the dynamics of such fields is controlled by the velocity properties
at their footpoints, which is here the interface between the turbulence
zone and the dynamo exterior.

\begin{figure}[t!]\begin{center}
\includegraphics[width=\columnwidth]{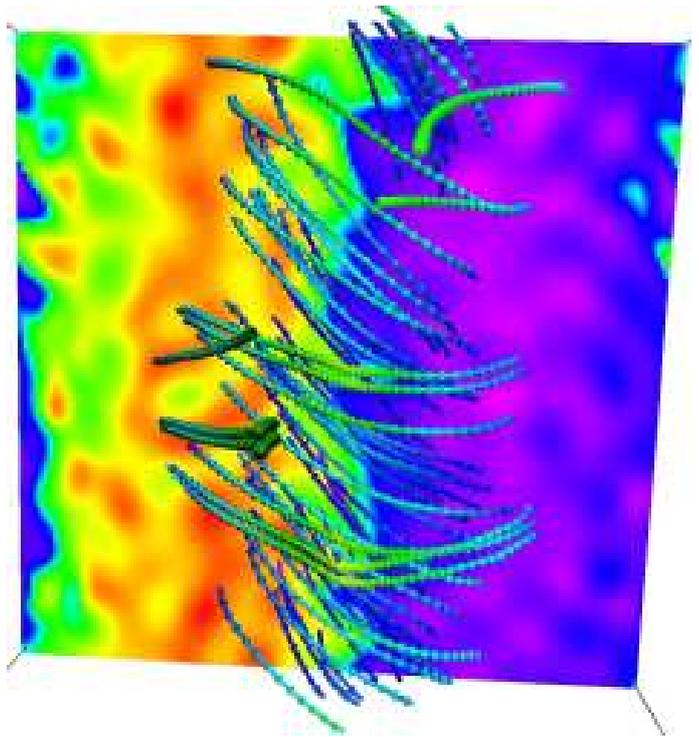}
\end{center}\caption[]{
Three-dimensional visualization of the magnetic field viewed from above.
The vertical magnetic field component is color-coded (yellow pointing
upward and blue pointing downward).
Note that the field lines form a left-handed spiral over the scale of
the domain, as expected for turbulence with positive helicity at
small scales.
}\label{fig_twist}
\end{figure}

\subsection{Force-free versus current free}
\label{ForceFree}

Already the straightforward inspection of magnetic field lines
viewed from the top suggests that the magnetic field is twisted
and forms a left-handed spiral; see \Fig{fig_twist}.
This is indeed the orientation expected for turbulence with positive
kinetic helicity, producing a  negative $\alpha$ effect and hence
magnetic spirals with negative helicity at the scale of the domain.

\begin{figure}[t!]\begin{center}
\includegraphics[width=\columnwidth]{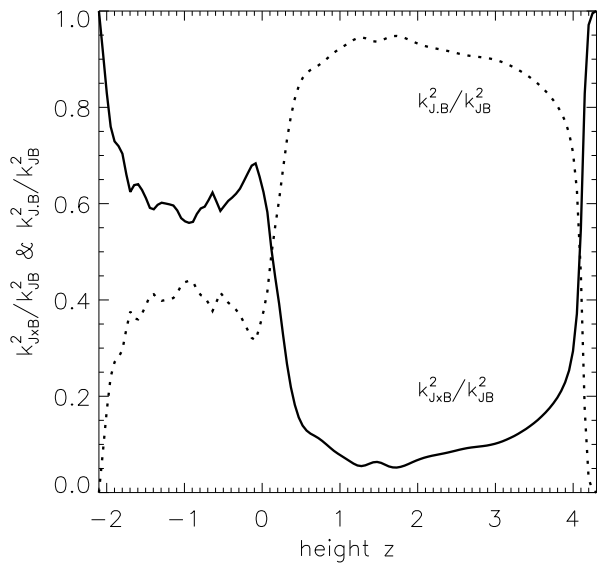}
\end{center}\caption[]{
Vertical dependence of $k^2_{J\times B}/k^2_{JB}$ and $k^2_{J\cdot B}/k^2_{JB}$.
Note the decline of the normalized Lorentz force from more than 60\%
in the turbulence zone to less than 10\% in the exterior.
}
\label{fig_forc}
\end{figure}

We expect the magnetic field in the dynamo exterior to be nearly
force free, i.e., we expect $\bra{\left(\JJ\times\BB\right)^2}_{\rm H}$ to be small compared
with $\bra{\BB^2}_{\rm H}\bra{\JJ^2}_{\rm H}$.
Here, $\bra{\cdot}_{\rm H}$ denotes an $xy$ average.
In order to characterize the degree to which this is the case,
we define the quantities
\begin{equation}
k^2_{J\times B}=\mu_0^2\frac{\bra{\left(\JJ\times\BB\right)^2}_{\rm H}}{\bra{\BB^4}_{\rm H}},\quad
k^2_{J\cdot B}=\mu_0^2\frac{\bra{\left(\JJ\cdot\BB\right)^2}_{\rm H}}{\bra{\BB^4}_{\rm H}},
\end{equation}
and note that
\begin{equation}
\frac{k^2_{J\times B}}{k^2_{JB}}+\frac{k^2_{J\cdot B}}{k^2_{JB}}= 1,
\end{equation}
with $k^2_{JB}=\mu_0^2{\bra{\JJ^2}_{\rm H}}/{\bra{\BB^2}_{\rm H}}$.
In \Fig{fig_forc} we show $k^2_{J\times B}$ and $k^2_{J\cdot B}$
as functions of $z$.
Given that $k^2_{J\times B}/k^2_{JB}$ has values below 0.1,
it is evident that the field is indeed nearly force free in the exterior.
In the turbulence zone, on the other hand, the Lorentz force is quite
significant.

\begin{figure}[t!]\begin{center}
\includegraphics[width=\columnwidth]{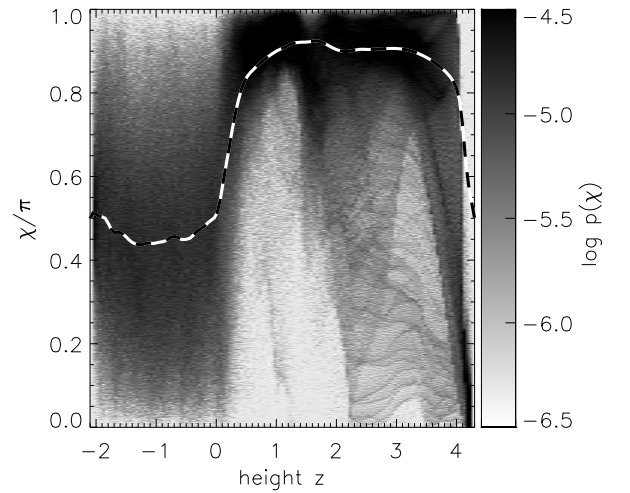}
\end{center}\caption[]{
Two-dimensional histogram of the distribution of angles,
$p(\chi)$, where $\chi=\arccos(\JJ\cdot\BB)$ is the angle between
$\JJ$ and $\BB$ at different heights.
$p(\chi)$ is normalized such that $\int p(\chi) d\chi = 1$.
The dashed line gives the location of the maximum position
of the distribution.
}
\label{fig_alpha}
\end{figure}

To prove the existence of a force-free structure, it is
convenient to calculate the angle $\chi$ between $\JJ$ and $\BB$.
We expect $\chi$ to be 0 or $\pi$ for an ideal force-free environment. 
In \Fig{fig_alpha} we show the distribution of values of $\chi$ plotted over the height $z$.
One sees that, in the exterior, the angle $\chi$ is close to $\pi$,
but it drops to $\pi/2$ at the very upper part, where the normal field
condition enforces that $\JJ$ and $\BB$ are at right angles to each other.

\begin{figure}[t!]\begin{center}
\includegraphics[width=\columnwidth]{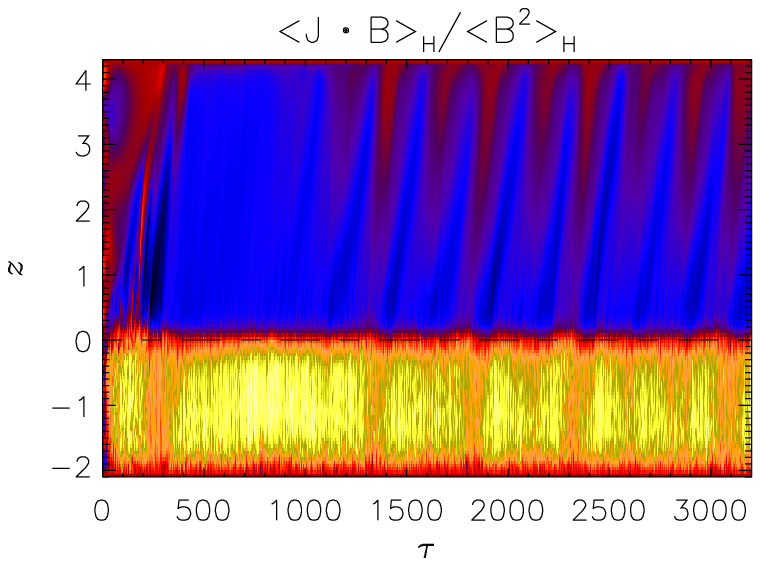}
\end{center}\caption[]{
Dependence of $\bra{\JJ\cdot\BB}_{\rm H}/\bra{\BB^2}_{\rm H}$
versus time $\tau$ and height $z$ for $L_z=6.4$ with $\Rm=3.4$ (Run~A).
}
\label{pjbm_cont}
\end{figure}

\begin{figure}[t!]\begin{center}
\includegraphics[width=\columnwidth]{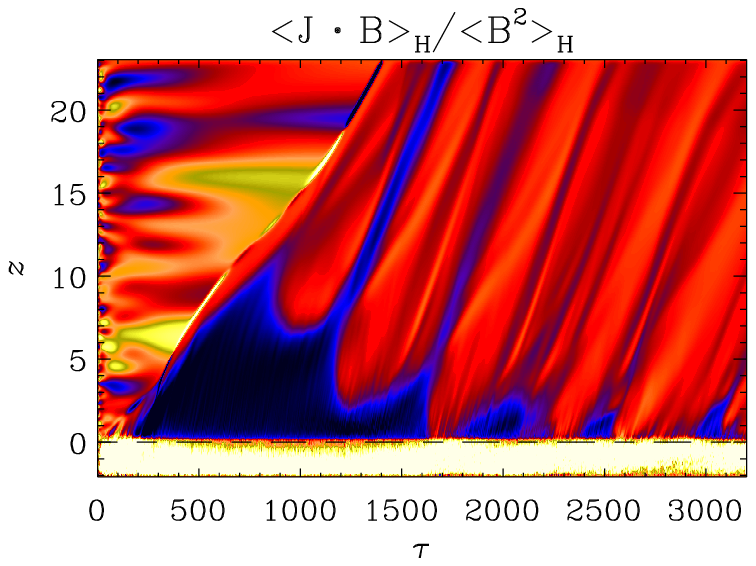}
\end{center}\caption[]{
Similar to \Fig{pjbm_cont}, but for $L_z=8\pi$ and $\Rm=6.7$
.
}
\label{pjbm_cont2}
\end{figure}

\begin{table*}[t!]\caption{
Summary of runs.
Run~A is for a cubic domain while Runs B1 to B7 are for taller domains
at different values of $\Rm$.
Note that the nondimensional interval length $\delta\tau$ as well as
the ejection speeds cannot be determined accurately for most of the runs,
but the values suggests that there is no systematic dependence on the
value of $\Rm$.
The column `$lm$' gives the $l$ and $m$ values of the leading mode
of the mean field in \Eq{MFdef} and the columns `direction' gives the
direction of propagation of this mean field, confirming that it is a
matter of chance.
}\vspace{12pt}\centerline{\begin{tabular}{lcccccccccc}
Run & $\Rey$ & $\Rm$  & $\Pm$ & $k_1w$ & $\Brms^2/\Beq^2$ & $\meanB_{i,{\rm rms}}^2/\Beq^2$ & $lm$ & direction & $\Delta\tau$  &  $V_{\rm eject}/\urms$ \\
\hline
A  & 3.4 &   3.4& 1 & 0.1 & 0.3-0.4 & 0.11 & 01 & $+y$ &  250 & 0.4488 \\
B1 & 6.7 &   6.7& 1 & 0.2 & 1.0-1.2 & 0.52 & 01 & $-y$ &  530 & 0.5464 \\
B2 & 6.7 &  13  & 2 & 0.2 & 0.9-1.2 & 0.45 & 10 & $+x$ &  570 & 0.628  \\
B3 & 6.7 &  67  &10 & 0.2 & 0.9-1.0 & 0.28 & 01 & $\pm y$ &  800 & 0.5463 \\
B4 & 6.7 & 133  &20 & 0.2 & 0.9     & 0.27 & 01 & $-y$ &  ?   &   ?    \\
B5 &15.0 &  15  & 1 & 0.2 & 0.7-1.0 & 0.29 & 10 & $+x$ &  370 & 0.5236 \\
\label{Summary}\end{tabular}}\end{table*}
In order to demonstrate that plasmoid ejection is a recurrent phenomenon,
it is convenient to look at the evolution of the ratio
$\bra{\JJ\cdot\BB}_{\rm H}/\bra{\BB^2}_{\rm H}$ versus $t$ and $z$.
This is done in \Fig{pjbm_cont} for $L_z=6.4$ and $\Rm=3.4$ (Run~A)
and in \Fig{pjbm_cont2} for $L_z=8\pi$ and $\Rm=6.7$ (Run~B1).
It turns out that in both cases the typical speed of plasmoid
ejecta is about 1/2 of the rms velocity of the turbulence in the
interior region.
However, the time interval $\delta\tau$ between plasmoid ejections
increases from $\approx250$ to $\approx570$ turnover times 
as we increase the kinetic and magnetic Reynolds numbers.
At higher magnetic Reynolds numbers, the length of the interval increases
to $\approx800$ turnover times.
A summary of all runs is given in \Tab{Summary}.
Here we also give the $l$ and $m$ values of the leading mode of the mean
field in \Eq{MFdef} and indicate explicitly whether the large-scale
magnetic field varies in the $x$ or the $y$ direction.
Indeed, both directions are possible, confirming that it is a matter of chance,
In most cases the magnetic field pattern shows a slow horizontal
migration, whose direction appears to be random.
The sign in the table indicates whether the wave migrates in the positive
or negative coordinate direction.

\subsection{Interpretation in terms of a mean-field model}

The magnetic field found in the simulations displays a clear large-scale
structure.
One may have expected that the magnetic field varies mainly in the
$z$ direction, because this is also the direction in which the
various profile functions vary.
However, this is not the case.
Instead, the main variation is in one of the horizontal directions
(see \Fig{cross-section}, where the field varies mainly in the $y$ direction).
The magnetic field does of course also vary in the $z$ direction,
but this happens without sign change in $\meanB_z$.
Above the surface at $z=0$, the field
gradually decays and retains only rather smooth variations.

In order to compare with dynamo theory, we now solve the usual
set of mean-field equations for the mean magnetic vector potential
$\meanAA$, where $\meanBB=\nab\times\meanAA$ is the mean magnetic field
and $\meanJJ=\nab\times\meanBB/\mu_0$ is the mean current density,
\EQ
{\partial\meanAA\over\partial t}=\alpha\meanBB-(\etat+\eta)\mu_0\meanJJ.
\label{MFdynamo}
\EN
We recall that $\eta=\const$ is the microscopic magnetic diffusivity,
which is not negligible, even though it is usually much smaller than
$\etat$.
We consider first the kinematic regime where $\alpha$ and $\etat$
are independent of $\meanBB$.
In order to account for the fact that there is no turbulence above
the turbulence zone, we adopt the profile $\theta_w(z)$ for $\alpha$
and $\etat$, i.e., we write
\EQ
\alpha(z)=\alpha_0\theta_w(z),\quad
\etat(z)=\etatz\theta_w(z),
\EN
where $\alpha_0$ and $\etatz$ are constants, and for $w$ we take the same
value as for the other profile functions used in the direct simulations.
The excitation condition for the dynamo can be quantified in terms
of a dynamo number that we define here as
\EQ
C_\alpha=\alpha_0/\etaTz k_1,
\EN
where $\etaTz=\etatz+\eta$ is the total magnetic diffusivity
and $k_1$ was defined in \Sec{TheModel} as the smallest
horizontal wavenumber that fits into the domain.
If the turbulence zone were homogeneous and periodic in the $z$ direction,
the critical value of $C_\alpha$ is unity,
but now the domain is open in the $z$ direction, so one expects the
dynamo to be harder to excite.
In the models presented below we therefore adopt the value
$C_\alpha=2.5$, which is also compatible with estimates of the critical
value from the simulations, if we write $\alpha\approx\urms/3$ and
$\etat\approx\urms/3\kf$.

Next, we consider the nonlinear regime by employing the dynamical quenching
model (Kleeorin \& Ruzmaikin 1982; Blackman \& Brandenburg 2002).
We assume that $\alpha=\alphaK+\alphaM$, where now $\alphaK(z)=\alpha_0\theta_w(z)$
is the kinetic $\alpha$ effect profile used earlier in the solution to the
kinematic equations, and $\alphaM$ is the solution to the dynamical quenching equation,
\EQ
{\partial \alpha_M \over\partial t}=
-2\etat\kf^2{\meanEMF\cdot\meanBB\over\Beq^2}
-2\eta\kf^2\alphaM,
\EN
where $\meanEMF=\alpha\meanBB-\etat\mu_0\meanJJ$ is the mean electromotive force.
Note that we have here ignored the possibility of magnetic helicity fluxes
that must become important at larger values of $\Rm$.

\begin{figure}[t!]\begin{center}
\includegraphics[width=\columnwidth]{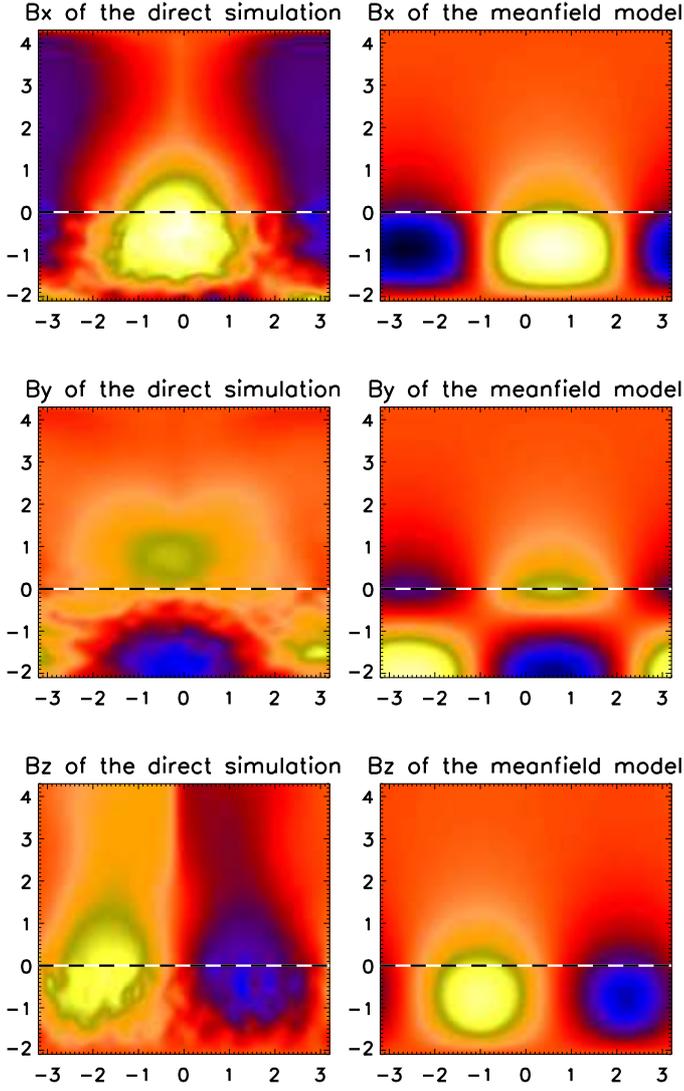}
\end{center}\caption[]{
Comparing the average in the $x$ direction of the magnetic
field with the 2D mean-field model. The structures are very similar.}
\label{fig_compare}
\end{figure}

\begin{figure}[t!]\begin{center}
\includegraphics[width=\columnwidth]{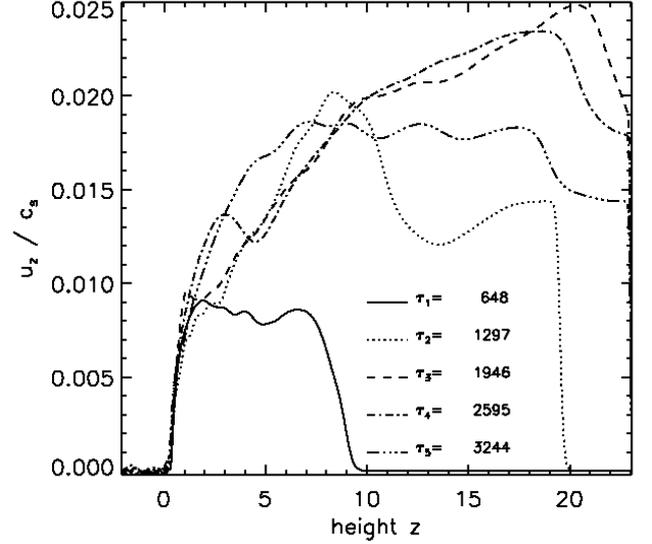}
\end{center}\caption[]{
Horizontally averaged rms velocity field as a function of height $z$
for 5 different times for a run with $L_z=8\pi$.
Note the development of a statistically steady state after about 1000
turnover times.
}
\label{fig_uzt}
\end{figure}

In \Fig{fig_compare} we compare the resulting magnetic field geometry
with that of the direct simulations.
In both cases the horizontal variation of the field is similar.
However, in the direct simulations the field extends more freely into the exterior.
This is probably caused by a vertical upward flow that appears to be driven by
the magnetic field.
In \Fig{fig_uzt} we see the vertical dependence of an upward flow,
which soon reaches a statistically steady state.

\section{Conclusions}

Our first results are promising in that the dynamics of the magnetic
field in the exterior is indeed found to mimic open boundary conditions
at the interface between the turbulence zone and the exterior at $z=0$.
In particular, it turns out that a twisted magnetic field generated
by a helical dynamo beneath the surface is able to produce flux emergence
in ways that are reminiscent of that found in the Sun.

Some of the important questions that still remain open include the
presence and magnitude of magnetic helicity fluxes.
In the present model we expect there to be diffusive magnetic helicity
fluxes associated with the vertical gradient of magnetic helicity density.
A related question concerns the dependence on the magnetic Reynolds number.
One expects magnetic helicity fluxes to become more important at large
values of $\Rm$.

One of the future extensions of this model includes the addition of shear.
In that case one might expect there to be strong magnetic helicity fluxes
associated with the Vishniac \& Cho (2001) mechanism that may transports magnetic
helicity along the lines of constant shear, although more recent
considerations now cast doubt on this possibility
(Hubbard \& Brandenburg 2010b).
One should also keep in mind that the magnetic field cannot really be
expected to be fully helical, as was assumed here in order to promote
large-scale dynamo action under relatively simple conditions.
Reducing the degree of helicity makes the dynamo harder to excite.
On the other hand, shear helps to lower the excitation conditions,
making it again feasible to obtain large-scale dynamo action even
at low relative helicity of the driving.
Another promising extension would be to move to a more global geometry,
including the effects of curvature and gravity.
This would allow for the emergence of a Parker-like wind that turns into
a supersonic flow at sufficiently large radii.
This would also facilitate the removal of magnetic field through the
sonic point.

\begin{acknowledgements}
We thank Jaime de la Cruz Rodriguez, Gustavo Guerrero, and
G\"oran Scharmer for discussions at the early stages of this work.
We acknowledge the allocation of computing resources provided by the
Swedish National Allocations Committee at the Center for
Parallel Computers at the Royal Institute of Technology in
Stockholm and the National Supercomputer Centers in Link\"oping.
This work was supported in part by
the European Research Council under the AstroDyn Research Project 227952
and the Swedish Research Council grant 621-2007-4064.
\end{acknowledgements}

\vfill\bigskip\noindent\tiny

\end{document}